\documentclass[twocolumn,showpacs]{revtex4}
\usepackage{graphicx}

\begin{document}

\title{Efficient generation of an isolated single-cycle attosecond pulse}
\author{Pengfei Lan, Peixiang Lu$\footnote{lupeixiang@mail.hust.edu.cn}$, Wei Cao, Xinlin Wang}
\affiliation{Wuhan National Laboratory for Optoelectronics and
School of Optoelectronics Science and Engineering, Huazhong
University of Science and Technology, Wuhan 430074, P. R. China}
\date{\empty}

\begin{abstract}
A new method for efficiently generating an isolated single-cycle
attosecond pulse is proposed. It is shown that the ultraviolet
(UV) attosecond pulse can be utilized as a robust tool to control
the dynamics of electron wave packets (EWPs). By adding a UV
attosecond pulse to an infrared (IR) few-cycle pulse at a proper
time, only one return of the EWP to the parent ion is selected to
effectively contribute to the harmonics, then an isolated
two-cycle 130-as pulse with a bandwidth of 45 eV is obtained.
After complementing the chirp, an isolated single-cycle attosecond
pulse with a duration less than 100 as seems achievable. In
addition, the contribution of the quantum trajectories can be
selected by adjusting the delay between the IR and UV fields.
Using this method, the harmonic and attosecond pulse yields are
efficiently enhanced in contrast to the scheme [G. Sansone {\it et
al.}, Science {\bf314}, 443 (2006)] using a few-cycle IR pulse in
combination with the polarization gating technique.
\end{abstract}
\maketitle

Attosecond (as) extreme ultraviolet pulses open the way to a new
regime of investigating and manipulating basic ultrafast
electronic processes in atoms and molecules with an unprecedented
precision
\cite{M.Hentschel,R.Kienberger,K.J.Schafer,P.Johnsson,P.M.Paul}.
Thus the generation of attosecond pulses has attracted spectacular
interest in recent years. So far high harmonic generation (HHG) is
the most promising way to produce attosecond pulses. In the
breakthrough work \cite{P.M.Paul}, a train of attosecond pulses
were produced with a multi-cycle laser pulse. However, the
straightforward attosecond metrology prefers an isolated
attosecond pulse \cite{I.P.Christov} and then much effort has been
paid to extract an isolated pulse from the attosecond pulse train.

HHG is well understood in terms of the ``recollision'' model
\cite{P.B.Corkum,M.Lewenstein}. In detail, the electron wave
packet (EWP) is first produced when the atom or molecule is
ionized. When the laser field drives the EWP back to the parent
ion, it interferes with the bound wavefunction, producing coherent
attosecond light bursts. If only a single return is possible, an
isolated attosecond pulse can be obtained. It was demonstrated
that the return probability of EWP to the parent ion depends
sensitively on the polarization of the driving field
\cite{M.Ivanov}. Using polarization gating technique
\cite{M.Ivanov}, the EWP only returns efficiently in a short time
and then an isolated attosecond pulse can be produced. An
alternative method to produce isolated attosecond pulses is using
a few-cycle laser pulse \cite{M.Hentschel,I.P.Christov}. It has
been shown that the highest harmonics (so-called cutoff) are only
generated at the peak of the few-cycle pulse. By filtering these
harmonics, an isolated attosecond pulse can be obtained. However,
the bandwidth of the harmonics in the cutoff is $\sim10$ eV, which
only supports to an attosecond pulse of about $250$ as
\cite{R.Kienberger}. Currently, an intense research is afoot to
push the pulse duration to even shorter time
\cite{K.T.Kim,Z.Chang,R.L.Martens,G.Sansone,I.J.Sola}. In the
resent work \cite{G.Sansone}, Sansone {\it et al.} produced an
isolated broadband attosecond pulse with a duration of $130$ as
using a few-cycle ($5$-fs) laser pulse with a modulated
polarization. However, there are some flaws for enhancing the
harmonic yield in this scheme, which limits the application of
this attosecond pulse. That is, the available energy of the
few-cycle driving laser pulse generated by the state-of-the-art
hollow fiber and chirped mirror scheme is usually less than 1 mJ.
Further, fractional energy of the few-cycle pulse is lost in the
polarization gating technique.

\begin{figure}
\begin{center}
\includegraphics[width=6cm,clip]{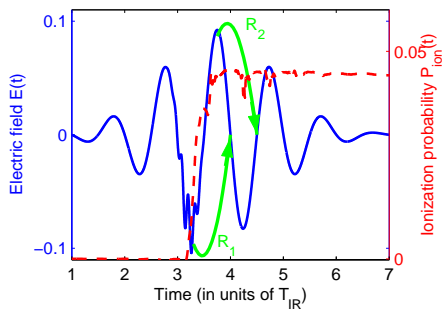}
\caption{\label{fig1} (color online) Schematic illustration (solid
line) of the harmonic generation process and the evolution of the
ionization probability (dashed line) in the combined field of an
IR few-cycle laser pulse and a UV attosecond pulse. The
intensities of the IR and UV fields are $3\times10^{14}$ W/cm$^2$
and $2\times10^{13}$W/cm$^2$, and the delay
$\omega_{IR}\tau_{delay}=-1.1\pi$.}
\end{center}
\end{figure}

In this work, we propose a new efficient method for the generation
of isolated broadband attosecond pulses using a few-cycle IR
driving laser pulse in combination with a UV attosecond
controlling pulse. We first analyse the HHG in the few-cycle IR
field alone. Due to the ultrashort duration, the EWP is dominantly
produced near the peak of the few-cycle pulse \cite{V.S.Yakovlev}.
The EWP produced at $t_1=3.2T_{IR}$ (marked by $R_1$ in Fig.
\ref{fig1}) is driven by the IR field at the highest amplitude,
and returns to the parent ion near $4T_{IR}$ with the energy of
about $3.2U_p$, where $T_{IR}$ and $U_p$ are the period of IR
field and ponderomotive energy, respectively. In the following
half cycle, the EWP is produced at $t_2=3.75T_{IR}$ (marked by
$R_2$ in Fig. \ref{fig1}) and returns to the parent ion at
$4.5T_{IR}$ with the energy of about $2.5U_p$. Therefore more than
one returns ($R_1$ and $R_2$) contribute to the lower harmonics
(so-called plateau), and the isolated attosecond pulse is only
achievable by filtering the harmonics with a bandwidth only
$0.7U_p$ ($\sim10$ eV), which prevents obtaining the broadband
isolated attosecond pulse with a shorter duration. If the
polarization of the few-cycle pulse is modulated, the contribution
from one return can be eliminated, then a broadband attosecond
pulse can be obtained \cite{Z.Chang,G.Sansone,I.J.Sola}. However,
the efficiency is limited as discussed in the previous paragraph.
In the present work, we propose an efficient method for producing
a broadband isolated attosecond pulse, this is achieved by
enhancing the contribution from one return with a UV pulse. Note
that the rapid advancement of HHG has made it available of a UV
pulse with an intensity of about $1\times10^{14}$ W/cm$^2$
\cite{Mashiko,Bandrauk}. Such attosecond pulses provide us a
robust tool to control EWPs \cite{P.Johnsson}, which have been
used in previous works \cite{Bandrauk,K.J.Schafer,A.D.Bandrauk}.
The schematic illustration of our method is shown in Fig.
\ref{fig1}. An attosecond UV controlling pulse is synthesized to
the few-cycle IR driving pulse at the time of $t_1$. Due to the
high photon energy and ultrashort duration of the UV pulse, the
production time and property of EWPs can be manipulated. Then, in
the combined field of an IR and a UV pulses, the EWP is only
effectively produced at $t_1$, returns to the parent ion at
$t=4T_{IR}$, and the contribution from the EWP produced at $t_2$
is much lower than that at $t_1$. Consequently, only the return
$R_1$ is selected to effectively contribute to the harmonics,
broadband harmonics and an isolated ultrashort attosecond pulse
will be obtained. It should be emphasized that the few-cycle IR
pulse also plays a vital role in this scheme, broadband isolated
attosecond pulse can not be produced with a multi-cycle pulse
\cite{A.D.Bandrauk,K.J.Schafer}. In contrast to the method using a
single few-cycle IR pulse in combination with the polarization
gating technique \cite{G.Sansone}, the energy loss of the IR field
is avoided in our scheme. Moreover, the UV field facilitates the
ionization, then the harmonic and attosecond pulse yields are
significantly enhanced.

To verify the above scheme, we perform a quantum simulation by
solving the time-dependent Schr$\ddot{o}$dinger equation, the
Hamiltonian is $H=-\partial^2/2\partial {x^2}+V(x)-xE(t)$ (in
atomic units), where $E(t)$ the electric field and $V(x)$ is the
atomic Coulomb potential. We choose the soft-core model within the
single-active electron approximation $V(x)=-1/\sqrt{\alpha+x^2}$,
where $\alpha$ is a smoothing parameter, routinely employed in
many studies of atomic ionization and HHG. Although our work is
not intended to model a specific experiment, in order to make the
discussion concrete, $\alpha$ is set to be $0.485$ to retrieve the
ionization energy of helium. And it has been shown that the
soft-core model well describes the real atom \cite{Q.Su}, we
expect our result is comparable to the HHG of helium. The electric
fields of the IR and UV pulses are given by
$E_j(t)=E_{0j}f_j(t-\tau_j)cos[\omega_j(t-\tau_j)], j=IR,UV.$
$E_{0j}$ and $\omega_j$ are the amplitudes and frequencies,
$f(t)_j$ and $\tau_j$ are envelopes and peak positions of the
pulses. Like Ref. \cite{Bandrauk}, a Gaussian envelope shape is
adopted and the delay of the UV pulse with respect to the IR pulse
is $\tau_{delay}=\tau_{UV}-\tau_{IR}$. For the IR driving field,
the intensity is $3\times10^{14}$ W/cm$^2$, the central wavelength
and pulse duration are $800$ nm and $5$ fs, respectively. For the
UV controlling field, the intensity is $2\times10^{13}$ W/cm$^2$,
the central wavelength is 100 nm, and pulse duration is $0.6$ fs
full width at half maximum, containing about 2 cycles. The
time-dependent Schr$\ddot{o}$dinger equation is solved with
split-operator method \cite{M.D.Feit}. The initial wavefunction is
chosen to be the field-free ground state which is obtained by
propagation in imaginary time. The ionization probability is given
by $P(t)=1-\sum_n<\phi_n|\Psi(t)>$ where the summation runs over
the bound states $\phi_n$. The time-dependent dipole acceleration
can be calculated with the Ehrenfest theorem \cite{K.Burnett}, and
then the harmonic spectrum is calculated by Fourier transforming
the dipole acceleration.

\begin{figure}
\begin{center}
\includegraphics[width=8.6cm,clip]{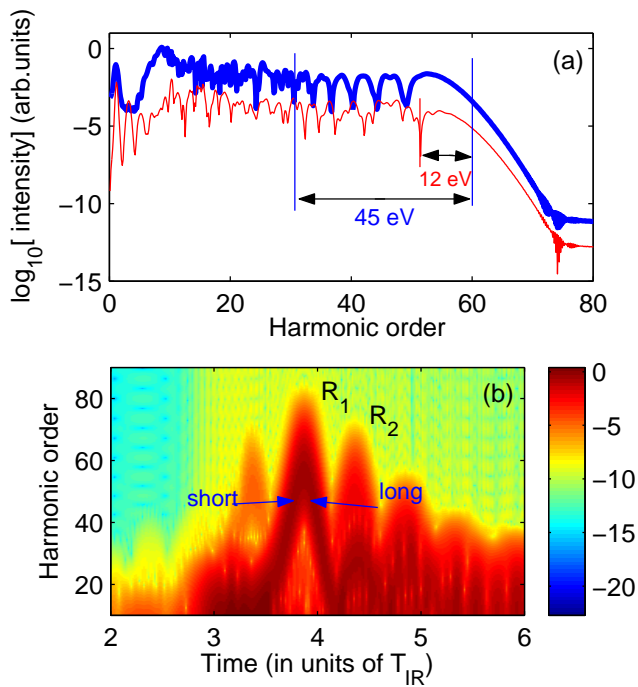}
\caption{\label{fig2} (color online) (a) The harmonic spectrum by
in the $5$-fs IR field alone (thin line) and in combination with a
$0.6$-fs UV field (bold line), respectively. (b)The time-frequency
characteristics of the harmonics in the combined field. The
parameters are the same with Fig. \ref{fig1}.}
\end{center}
\end{figure}

The dashed line in Fig. \ref{fig1} shows the evolution of the
ionization probability in the IR driving pulse combined with an
attosecond UV controlling pulse at $t_1=3.2T_{IR}$. One can see
that the ionization probability increases steeply at $t_1$ and
varies slightly at other time. This indicates that the EWP is only
effectively produced at $t_1$ which agrees with our proposal. The
ionization probability is much higher than that in the IR or UV
field alone, which will significantly enhance the harmonic and
attosecond pulse yields. This is an effect of cross correlation of
the two pulses, which can be explained as follows. The 0.6-fs UV
pulse spans a bandwidth of several eV, it can promote electronic
transition from ground state to the first excited state (with the
eigenenergy of -$0.32$). And on the other hand, IR laser photons
also can assist the electronic transition. Thus, the electron in
the ground state is first pumped to the excited state, and then is
up-shifted and easily ionized by the IR driving field because of
the low binding energy. This explanation agree well with Ref.
\cite{K.Ishikawa}. Also, Paul {\it et al.} \cite{Paul} observed
significant HHG enhancement with the prepared excited atom
recently. From this viewpoint, our result also can be explained
that an excited atom is prepared by the UV pulse in combination
with the IR pulse, and so the ionization and HHG is significantly
enhanced. Note that this process is different from Ref.
\cite{K.J.Schafer}, where the photon energy of the UV pulse is
higher than the ionization energy and the electron is dominantly
ionized by absorbing one UV photon. Fig. \ref{fig2}(a) shows the
harmonic spectrum by an IR few-cycle laser pulse alone (the thin
line). One can see that the overall spectral structure is
irregular for the harmonics bellow $52\omega_{IR}$, and is smooth
for the harmonics in the cutoff. The isolated attosecond pulse is
only achievable by selecting the highest harmonics with a
bandwidth of about $12$ eV. While in the combined field of an IR
and UV pulses, the harmonic spectrum shows a different structure
in contrast to that in the IR pulse alone. One can see that (the
bold line) the harmonic spectrum is smooth and regularly modulated
from $31\omega_{IR}$. In addition, the harmonic yield is about two
orders of magnitude higher than that in the IR driving field
alone.

To understand the spectral structure shown in Fig. \ref{fig2}(a),
the time-frequency image of the harmonic spectrum
\cite{P.Antoine2} in the combined field is shown in Fig.
\ref{fig2}(b). One can see that there are two dominant returns
contributing to the HHG near the peak of the driving field. For
the first one, the harmonics are emitted at $t=4T_{IR}$, and the
highest energy is about $60\omega_{IR}$ which is approximately
equal to $I_p+3.2U_p$. All these features consist with the
characteristics of the return $R_1$ and indicate that these
harmonics are generated by the EWP produced at $3.2T_{IR}$. For
the second return, the harmonics are emitted at $t=4.5T_{IR}$,
which correspond to the return $R_2$. However, in contrast to the
HHG in the IR field alone (see Fig. 1 of Ref.
\cite{V.S.Yakovlev}), one can see from Fig. \ref{fig2}(b) that the
first return ($R_1$) is highly enhanced in the combined field.
Thereby only the return $R_1$ is selected to effectively
contribute to the harmonics, and thus the harmonic spectrum
becomes more regular and smooth as shown in Fig. \ref{fig2}(a).
For the harmonics contributed from $R_1$, it is shown in Fig.
\ref{fig2}(b) that there are two emission times corresponding to
the same harmonic in the half optical cycle of the IR field. They
correspond to the short and long quantum trajectories
\cite{P.B.Corkum,M.Lewenstein}. The interference of these two
quantum trajectories leads to an evident modulation in the
harmonic spectrum. The spatial analogy of this phenomena is
Young's two-slit experiment. This interference will be weak when
eliminating one quantum trajectory, then the modulation is weak
also and the harmonic spectrum will become a supercontinuum
\cite{Z.Chang}.

\begin{figure}
\begin{center}
\includegraphics[width=6cm,clip]{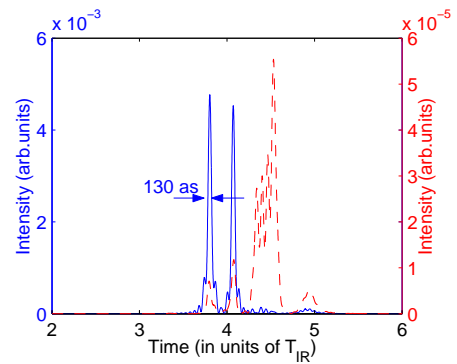}
\caption{\label{fig3} (color online) Temporal profiles of the
attosecond pulses by the combined field (solid line) and IR field
alone (dashed line), respectively. The parameters are the same
with Fig. \ref{fig1}.}
\end{center}
\end{figure}

As shown in Fig. \ref{fig2}, the spectral structure in the
combined field is different from that in the IR driving field
alone, which will produce different attosecond pulses. Fig.
\ref{fig3} shows the temporal profiles of the attosecond pulses
generated in the IR driving field alone (dashed line) and combined
field (solid line) by superposing the harmonics from
$31\omega_{IR}$ to $50\omega_{IR}$, respectively. In the IR field
alone, a train of chaotic attosecond pulses are generated (see the
dashed line). It is because both the return R$_1$ and R$_2$
contribute to the harmonics in the plateau, and different returns
take different trajectories and return times. Hence these high
harmonics are not synchronized, a train of chaotic attosecond
pulses are generated and it is hard to extract an isolated
attosecond pulse from this pulse train. While in the combined
field, a double-peak attosecond pulse is generated near
$t=4.0T_{IR}$, which is originated from the return of $R_1$. In
addition, the magnitude of the double-peak attosecond pulse is
enhanced by about two orders in the combined field. Note that
double-peak structure corresponds to the short and long
trajectories [Fig. \ref{fig2}(b)], respectively. After selecting
one trajectory, the high harmonics will be well synchronized and
an isolated broadband attosecond pulse will be obtained.
Generally, this can be achieved utilizing the phase-match
condition. It has been demonstrated that the short trajectory can
be macroscopically selected by adjusting the position of the laser
focus with respect to the gas jet \cite{G.Sansone,P.Antoine}.
Alternatively, the long trajectory leads to a spatially divergent
radiation \cite{M.Bellini}, and can be macroscopically eliminated
by adding a small aperture after the harmonic generation cell
\cite{R.L.Martens}. With these techniques, only the short
trajectory is selected and an isolated broadband attosecond pulse
will be produced \cite{K.T.Kim,Z.Chang,G.Sansone}.

\begin{figure}
\begin{center}
\includegraphics[width=8cm,clip]{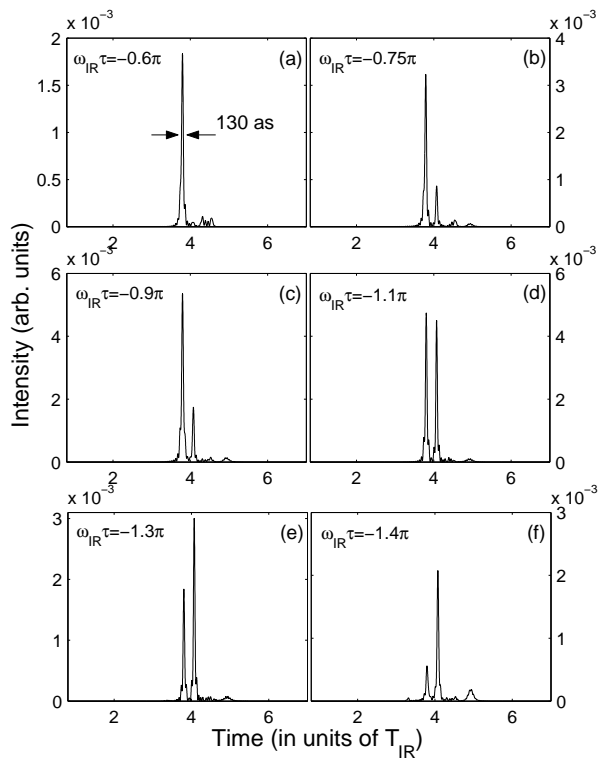}
\caption{\label{fig4} Temporal profiles of the attosecond pulses
generated at different delays between the IR and UV fields. An
isolated broadband attosecond pulse is generated as shown in Fig.
\ref{fig4}(a).}
\end{center}
\end{figure}
Here we show that the quantum trajectory also can be selected by
adjusting the delay between the IR and UV fields, then the high
harmonics can be well synchronized and an isolated broadband pulse
is generated [see Fig. \ref{fig4}(a)]. Fig. \ref{fig4} shows the
temporal profiles of the attosecond pulses generated in the
combined field for various delays. One can see that both the
intensity and the profile of the attosecond pulses vary
significantly by adjusting the delay. At
$\omega_{IR}\tau_{delay}=-1.1\pi$ [see Fig. \ref{fig4}(d)], both
the short and long trajectories contribute to the harmonics, and
two attosecond bursts with equal intensities are generated. With
decreasing the delay [Fig. \ref{fig4}(d)-(a)], the contribution
from the long trajectory becomes weak, and the intensity of the
post-pulse is lower. With increasing the delay [Fig.
\ref{fig4}(d)-(f)], the contribution from the short trajectory
becomes weak, and the intensity of the pre-pulse is lower. The
most fascinating result shown in Fig. \ref{fig4}(a) is that an
isolated broadband attosecond pulse is obtained at
$\omega_{IR}\tau_{delay}=-0.6\pi$, only accompanied by a minor
post-pulse with the magnitude of about $5\%$ of the main pulse
which can be ignored. The duration of this isolated attosecond
pulse is only $130$ as, containing only $1.95$ optical cycles of
the central harmonic frequency ($40\omega_{IR}$). In addition, it
is worth emphasizing that the bandwidth of the attosecond pulse is
broadened to $45$ eV. In the Fourier limit, a 90-as pulse will be
obtained. However, there is an intrinsic chirp in these harmonics
[see Fig. \ref{fig2}(b)], which prevents producing the sub-100 as
pulse. This problem can be overcome by using a proper material to
compensate the intrinsic chirp \cite{R.L.Martens,G.Sansone}. Then
an isolated single-cycle attosecond pulse with a duration less
than 100 as will be obtained. The efficient generation of isolated
single-cycle attosecond pulses may pave the way to investigate and
control the electron dynamics in atoms and molecules with a higher
precision. In addition, the few-cycle isolated attosecond pulse
can serve as a tool to study the electron dynamics in atoms and
molecules when processes are triggered and controlled by the
electric field of the attosecond pulse rather than by the
intensity profile. As in the case of few-cycle IR pulses, we
expect the possibility to influence and modify the ionization
mechanism of electrons upon changing the carrier-envelope phase of
the attosecond pulse in the extreme UV regime.

Few-cycle Ti:sapphire laser pulses are currently available in
quite a few laboratories around the world, which facilitates the
implementation of our scheme. In experiments, we can split a
$5$-fs Ti:sapphire laser pulse to two beams. Focusing one laser
beam to a gas jet (e.g. xenon), the 0.6-fs controlling UV pulse
can be produced by synthesizing the high harmonics like Refs.
\cite{M.Hentschel,R.Kienberger}, and then the generated UV pulse
is combined with the other laser beam. Following, the combined
field is focused to helium gas, and an isolated broadband $130$-as
pulse will be generated. Note that the delay between the UV pulse
and IR laser pulse should be adjust (see Fig. \ref{fig4}).
Establishing a dedicated delay shift can be done by introducing
small amounts of glass in experiments.

In conclusion, a new method for the efficient generation of an
isolated single-cycle attosecond pulse using a few-cycle IR
driving pulse in combination with a UV attosecond controlling
pulse has been proposed. It is shown that the EWP dynamics can be
controlled by the UV attosecond pulse. Then only one return of the
EWP to the parent ion is selected, and an isolated broadband
attosecond pulse of $45$ eV is generated. In addition, the
contribution of the short and long trajectories can be selected by
adjusting the delay between the IR and UV fields. The results show
that an isolated two-cycle attosecond pulse with a duration of
$130$ as is obtained. After compensating the chirp, it is expected
to produce an isolated single-cycle attosecond pulse with a
duration less than 100 as. For our scheme, the energy of the
few-cycle IR field is efficiently used, and the UV field
facilitates the ionization, thus the harmonic and attosecond pulse
yields are significantly enhanced.

This work was supported by the National Natural Science Foundation
of China under grant No. 10574050, the Specialized Research Fund
for the Doctoral Program of Higher Education of China under Grant
No. 20040487023 and the National Key Basic Research Special
Foundation under grant No. 2006CB806006.


\begin{thebibliography}{99}
\bibitem{M.Hentschel}
M. Hentschel {\it et al.}, Nature {\bf 414}, 509 (2001).
\bibitem{R.Kienberger}
R. Kienberger {\it et al.}, Nature (London), {\bf 427}, 817
(2004).
\bibitem{K.J.Schafer}
K. J. Schafer {\it et al.}, Phys. Rev. Lett. {\bf 92}, 023003
(2004).
\bibitem{P.Johnsson}
P. Johnsson {\it et al.}, Phys. Rev. Lett. {\bf 95}, 013001
(2005).

\bibitem{P.M.Paul}
P. M. Paul {\it et al.}, Science {\bf 292}, 1689 (2001).

\bibitem{I.P.Christov}
I. P. Christov, M. M. Murnane and H. C. Kapteyn, Phys. Rev. Lett.
{\bf 78}, 1251 (1997).

\bibitem{P.B.Corkum}
P. B. Corkum, Phys. Rev. Lett. {\bf 71}, 1994 (1993).
\bibitem{M.Lewenstein}
M. Lewenstein {\it et al.}, Phys. Rev. A {\bf 49}, 2117 (1994).
\bibitem{M.Ivanov}
M. Ivanov, P. B. Corkum, T. Zuo, and A. Bandrauk, Phys. Rev. Lett.
{\bf74}, 2933 (1995); P. B. Corkum, N. H. Burnett, and M. Ivanov,
Opt. Lett. {\bf19}, 1870 (1994).

\bibitem{R.L.Martens}
R. L$\acute{o}$pez-Martens {\it et al.}, Phys. Rev. Lett. {\bf
94}, 033001 (2005).
\bibitem{K.T.Kim}
K. T. Kim, {\it et al.} Phys. Rev. A {\bf69}, 051805(R) (2004); Z.
Chang, Phys. Rev. A {\bf71}, 023813 (2005).
\bibitem{Z.Chang}
Z. Chang, Phys. Rev. A {\bf 70}, 043802 (2004); V. Strelkov {\it
et al.}, J. Phys. B {\bf 38}, L161 (2005).
\bibitem{G.Sansone}
G. Sansone {\it et al.}, Science {\bf314}, 443 (2006).

\bibitem{I.J.Sola}
I. J. Sola {\it et al.}, Nature Physics {\bf 2}, 319 (2006).
\bibitem{V.S.Yakovlev}
V. S. Yakovlev, and A. Scrinzi, Phys. Rev. Lett. {\bf 91}, 153901
(2003).

\bibitem{Mashiko}
H. Mashiko, A. Suda, K. Midorikawa, Opt. Lett. {\bf29}, 1927
(2004).
\bibitem{Bandrauk}
A. D. Bandrauk, S. Chelkowski, N. H. Shon, Phys. Rev. Lett. {\bf
89}, 283903 (2002).

\bibitem{A.D.Bandrauk}
A. D. Bandrauk and N. H. Shon, Phys. Rev. A {\bf66}, 031401(R)
(2002).
\bibitem{Q.Su}
Q. Su, J. H. Eberly, Phys. Rev. A {\bf 44}, 5997 (1991); S. C.
Rae, X. Chen, K. Burnett, Phys. Rev. A {\bf 50}, 1946 (1994).
\bibitem{M.D.Feit}
M. D. Feit, J. A. Fleck, Jr, and A. Steiger, J. Comput. Phys.
{\bf47}, 412 (1982).
\bibitem{K.Burnett}
K. Burnett, V. C. Reed, J. Cooper, and P. L. Knight, Phys. Rev. A
{\bf45}, 3347 (1992).
\bibitem{K.Ishikawa}
K. Ishikawa, Phys. Rev. Lett. {\bf 91}, 043002 (2003).
\bibitem{Paul}
P. M. Paul, {\it et al.}, Phys. Rev. Lett. {\bf 94}, 113906
(2005).

\bibitem{P.Antoine2}
P. Antoine, B. Piraux and A. Maquet, Phys. Rev. A {\bf 51}, R1750
(1995).

\bibitem{P.Antoine}
P. Antoine, A. L'Huillier, and M. Lewenstein, Phys. Rev. Lett.
{\bf 77}, 1234 (1996).

\bibitem{M.Bellini}
M. Bellini {\it et al.}, Phys. Rev. Lett. {\bf 81}, 297 (1998).



\end{thebibliography}
\end{document}